\documentclass[preprint,5p,twocolumn]{elsarticle}

\usepackage[colorlinks = true,linkcolor = blue,urlcolor  = blue, citecolor = blue]{hyperref}
\usepackage{graphicx}
\usepackage{siunitx}
\usepackage{subcaption}
\usepackage{booktabs}
\usepackage{amsmath}
\usepackage{textcomp}
\usepackage{footnote}

\usepackage{svg}

\usepackage{amssymb}
\usepackage[version=4]{mhchem}

\usepackage[capitalise]{cleveref}

\journal{Nuclear Fusion}

\begin{document}

\begin{frontmatter}

\title{Physics-informed tritium fuel cycle modelling workflow for fusion reactors}

\cortext[cor1]{Corresponding author}

\author[MIT]{Rémi Delaporte-Mathurin\corref{cor1}}%
\ead{remidm@mit.edu}

\author[UKAEA]{Ross MacDonald}
\author[MIT]{James Dark}
\author[milan]{Milan Rother}
\author[MIT]{Tasnim Zulfiqar}
\author[MIT]{Kevin B. Woller}

\address[MIT]{Plasma Science and Fusion Center, Massachusetts Institute of Technology, Cambridge, MA 02139, USA}
\address[UKAEA]{United Kingdom Atomic Energy Authority, Culham Campus, Abingdon, Oxfordshire, OX14 3DB, UK}
\address[milan]{Independent Researcher}
\begin{abstract}

In this work, we present a multi-fidelity, physics-informed framework for tritium fuel cycle modelling based on the open-source PathSim/PathView platform.

Three complementary modelling approaches are demonstrated within a unified dynamic simulation environment. First, a zero-dimensional residence time model is used to reproduce the fuel cycle behaviour of an ARC-class fusion power plant, providing a baseline system-level description.
Second, an intermediate-fidelity component model based on coupled one-dimensional ordinary differential equations is developed to describe tritium mass transfer in a liquid metal bubble column reactor and validated against published literature before integration into the full fuel cycle.
Finally, high-fidelity multi-dimensional tritium transport models implemented using the finite element code FESTIM are coupled directly to the system model, enabling the inclusion of multi-dimensional effects, material interfaces, and complex transport phenomena.

This work demonstrates how fuel cycle components of varying physical fidelity can be combined consistently within a single, open-source framework.
The proposed approach enables more physically grounded fuel cycle analyses while retaining the flexibility required for system-level studies and provides a foundation for future integration with neutronics, fluid dynamics, and surrogate modelling tools.

\end{abstract}



\end{frontmatter}


\section{Introduction}

Fusion energy offers the potential for an abundant, low-carbon, and secure energy source, yet achieving tritium self-sufficiency remains one of its most critical challenges \cite{ferry_fusion_2024, delaporte-mathurin_advancing_2024}.
Tritium, the primary fuel for deuterium-tritium fusion reactions, is radioactive and scarce in nature, requiring that future power plants breed it in situ through reactions between neutrons and lithium within breeding blankets.
Ensuring a tritium breeding ratio (TBR) greater than unity is a necessary but not sufficient condition for tritium self-sufficiency of a fusion power plant fleet.
The efficiency and dynamic performance of the tritium fuel cycle —from breeding, extraction, and processing to reinjection into the plasma — strongly influence key metrics such as the tritium doubling time, the time a plant requires to reproduce its initial tritium inventory and start up additional reactors.

To date, several studies have used simplified, system-level approaches to analyse tritium fuel cycles. The residence time method, introduced by Abdou \textit{et al} \cite{abdou_deuterium-tritium_1986} and later developed in more details \cite{hattab_analysis_2025, simon_moose-based_2025, meschini_impact_2025, meschini_modeling_2023, baiquan_mean_2001}, remains a cornerstone for early-stage fuel cycle design.
This method describes each subsystem of the fuel cycle (such as breeding blankets, extraction systems, and isotope separation) in terms of its characteristic \textit{tritium residence time} and loss fraction.
Such zero-dimensional models allow designers to estimate system inventories, evaluate the feasibility of achieving tritium self-sufficiency under specific TBR and startup inventory constraints, and identify critical components requiring faster processing or reduced inventory.

However, the residence time approach has inherent limitations.
It relies on empirical or assumed values for residence times and does not capture the underlying physics governing tritium transport, retention, and exchange in individual subsystems.
Many key components (for example, tritium extraction systems or permeators) depend on parameters such as mass transfer coefficients, diffusion rates, and interface kinetics that cannot be accurately represented by a single effective residence time \cite{teichmann_fusion_2025, humrickhouse_tritium_2018}.
Moreover, these simplified models cannot account for spatially varying conditions, nonlinear coupling between processes, or time-dependent transients relevant to realistic plant operation.

To address these challenges, we propose a physics-informed framework for tritium fuel cycle modelling that integrates first-principles models directly into system-level simulations.
In this approach, each component can be represented by an appropriate level of physical fidelity: from simplified ordinary differential equations (ODEs) to detailed finite-element models, such as those produced with FESTIM \cite{dark_festim_2025, delaporte-mathurin_festim_2024}.
This multi-fidelity concept allows combining high-level system connectivity with physically grounded component behaviour, enabling both conceptual design and predictive analysis.

This work uses PathSim and PathView for fuel cycle simulations.

PathSim \cite{rother_pathsim_2025} is a Python based open source software package for dynamical system modelling and transient simulation. Its modular and distributed approach to system simulation provides the seamless integration of external simulation tools such as FESTIM \cite{dark_festim_2025, delaporte-mathurin_festim_2024} encapsulated as blocks, together with blocks from the built in library or custom blocks in a block diagram. PathSim offers a range of built in blocks from different disciplines such as PID controllers, spectrum analysers, or simple amplifiers and integrators that can be combined to form complex interconnected models of hundreds or thousands of blocks. The key benefit of this block based approach is the modularity of the resulting models. Blocks can easily be interchanged by high fidelity full physics models or cheap surrogate models depending on the demands of the simulation. Hierarchical modelling is supported through subsystems to abstract away the lower level complexity for high level design and modelling decisions.

This enables system level coupling of all the different components with their own physics that occur in fusion fuel cycles through a unified framework.
PathSim takes the role of the coordinator.
It arranges all the component models with varying fidelity -- from static models, to residence time models, to physics based models, to full FEM model integrations with i.e. FESTIM -- in the global timestepping loop.



However, building models solely through Python scripting can be cumbersome and error-prone beyond a certain scale and complexity.
Developing an intuition and interpretation for system layout as well as visual inspection are important aspects of the system development process.
This is where PathView comes in.
It bridges the user experience gap by providing a web-based graphical interface for PathSim using JavaScript/React.
Through it, users can quickly and intuitively build, modify and evaluate models without keeping the whole abstract system topology in their head.


PathView integrates all of PathSim's functionality including solver selection, simulation settings, extensibility through custom blocks and toolboxes, defining events and python objects in an integrated Python code editor.
On top of that, PathView offers post processing and interactive visualisation of the simulation results through a Plotly integration. 



For integration with external workflows and model exchange for collaboration, PathView can generate Python code from the graph. These stand-alone scripts can be executed independent of the graphical interface through the PathSim API.

\section{Residence time method: ARC fuel cycle}

\begin{figure*}[h!]
     \centering
     \begin{subfigure}[b]{\linewidth}
         \centering
         \includegraphics[width=0.7\linewidth]{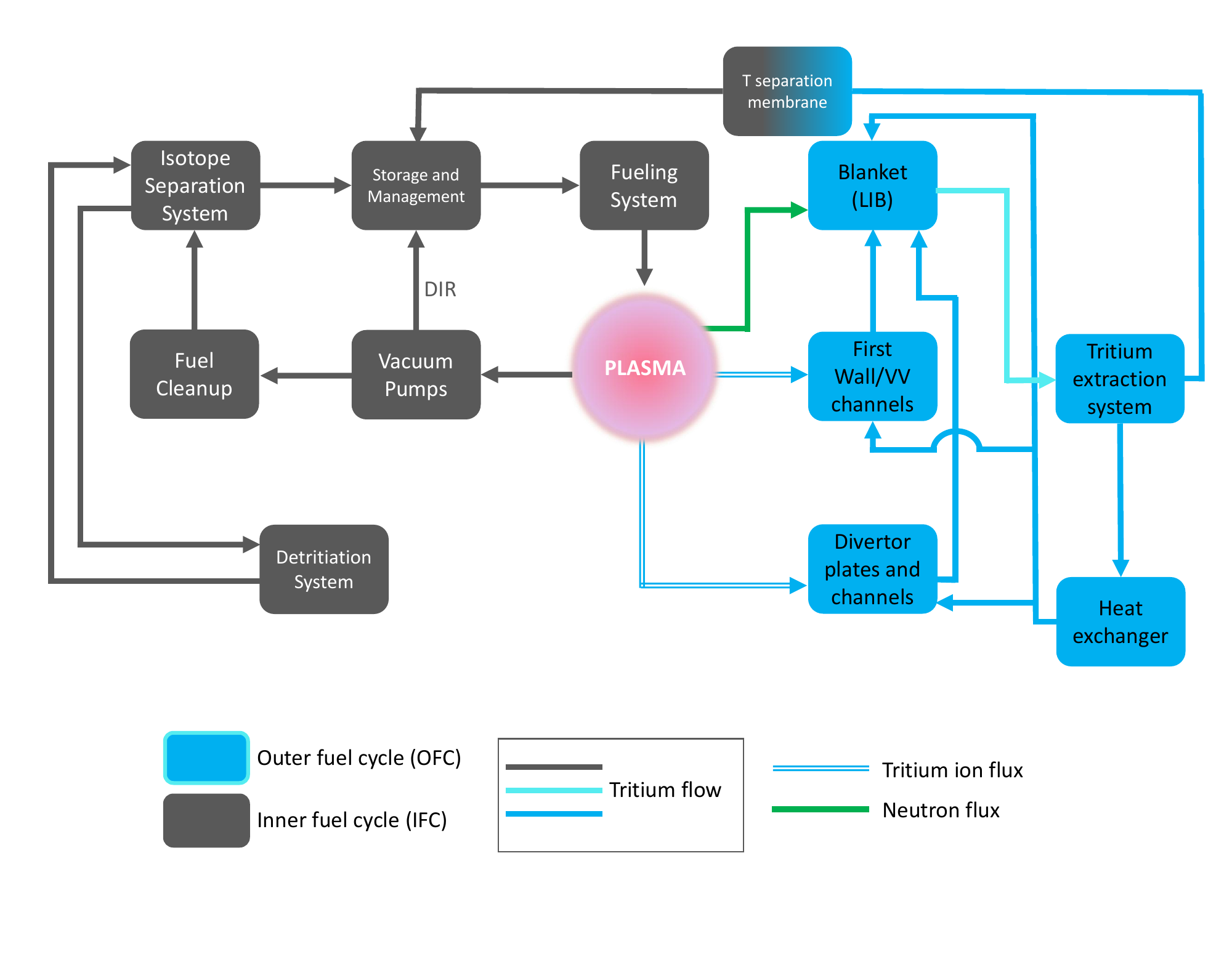}
             \caption{ARC-class fuel cycle diagram \cite{meschini_modeling_2023}.}
    \label{fig:arc model sketch}
     \end{subfigure}
       \hfill
     \begin{subfigure}{0.48\linewidth}
         \centering
         \includegraphics[width=\linewidth]{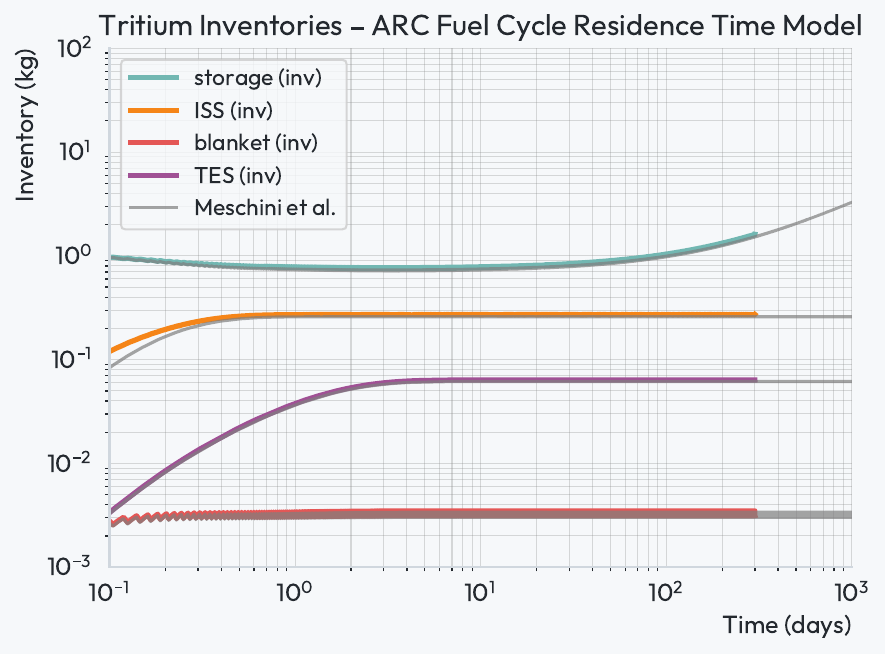}
             \caption{Comparison between Meschini \textit{et al} \cite{meschini_modeling_2023} (grey lines) and our model (coloured lines).}
    \label{fig:meschini comparison}
     \end{subfigure}
       \hfill
      \begin{subfigure}{0.48\linewidth}
         \centering
      
    \includegraphics[width=\linewidth]{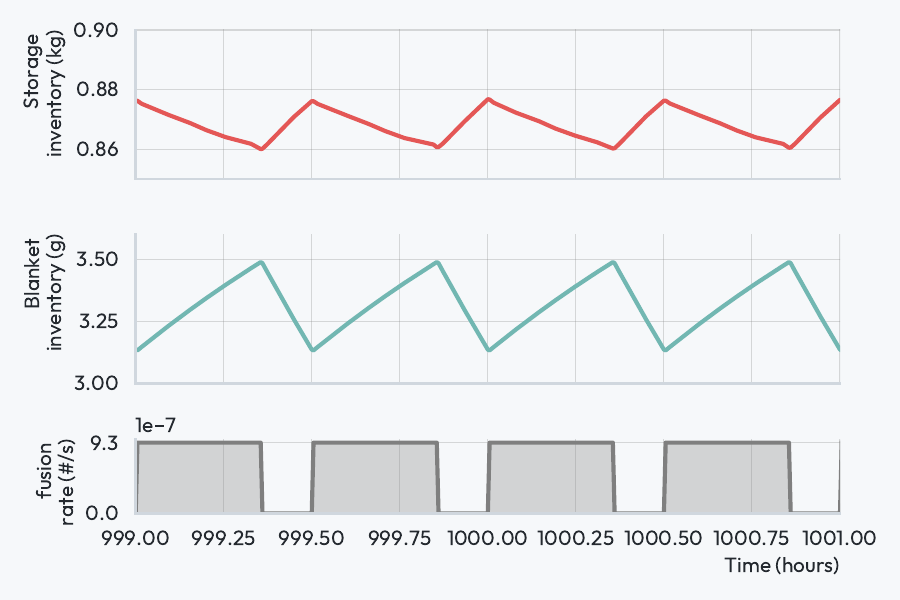}
    \caption{Temporal evolution of the storage inventory (top), blanket inventory (middle), and fusion rate (bottom).}
     \end{subfigure}
\end{figure*}

In this section, we start by illustrating the low-detail residence time method applied to an ARC-class fusion power plant fuel cycle (see Figure \ref{fig:arc model sketch}), following the implementation from Meschini \textit{et al} \cite{meschini_modeling_2023}.

In the residence time method, the tritium inventory in component $i$ is governed by the following equation:
\begin{equation}
    \frac{dI_i}{dt}=\sum_j F_{\mathrm{in}, j}-(1+\epsilon_i)\left(\frac{I_i}{\tau_i}\right)-\lambda I_i + S_i
    \label{eq:RTM}
\end{equation}
where $I_i$ is the tritium inventory in \si{kg}, $\sum_j F_{\mathrm{in}, j}$ is the total tritium mass flux in \si{kg/s} towards the component, $\tau_i$ is the residence time in \si{s}, $\epsilon_i$ is the non-radioactive losses fraction, $\lambda$ is tritium's decay constant in \si{s^{-1}}, and $S_i$ is the volumetric source term in \si{kg/s}.

This method is extremely useful to perform preliminary design specifications but is rapidly limited for more advanced design.

A first example is the influence of trapping.
Meschini \textit{et al} further improved this model by including thermally-activated tritium trapping as well as the influence of neutron-damage induced traps \cite{meschini_impact_2025, dark_modelling_2024}.
Results indicated that including trapping could seriously affect the overall fuel cycle efficiency.
However, because of the low-detail zero-dimensional nature of the residence time method, it required some simplifications and assumptions regarding the spatial dependence of 
trapping.

The ARC fuel cycle model in Meschini \textit{et al} was reimplemented in PathView using the standard case parametrisation - neglecting the non-radioactive losses (ie. $\epsilon_i = 0$) and radioactive decay (ie. $\lambda = 0$) for simplicity.
The results are in very good agreement (see Figure \ref{fig:meschini comparison}).
The slight differences in ISS inventory at the beginning of the simulation are attributed to the fact that losses (radioactive and non-radioactive) are neglected in the current study for simplification purposes.

In this model, the plasma source is a periodic step function based on an availability factor.
At $t=0$, all components have an initial tritium inventory of zero, except for the storage system, which has a startup inventory.
Before $t\approx \SI{10}{days}$, the storage inventory is depleted until the tritium bred in the outer fuel cycle (ie. Blanket, TES, etc.) flows back to the storage system.
At this time, the system reaches a pseudo-steady state and the storage inventory increases while other systems have a stable inventory.

The following sections will show examples of physics-informed fuel cycle components models in PathView/PathSim.

\section{1D Bubble Column Reactor Model}
A powerful capability of PathSim / PathView is the ability to create custom function blocks. This enables complex physics-informed models to be easily integrated into a wider dynamic system model. 
In this section we present an example of this with a Gas-Liquid Contactor (GLC) Bubble Column Reactor (BCR) model of Tritium extraction from liquid Lithium-Lead eutectic alloy (LiPb).

To increase the rate of mass transfer from a liquid to a gas, a GLC should maximise the interfacial area over which the flux of tritium mass transfer from the liquid to gas phase, $J_{T}$, acts. A BCR achieves this by bubbling gas through a column of liquid, which results in a large specific interfacial area $a$ between the phases. 

There are various designs of BCR; for example, the two phases can flow in concurrent or countercurrent directions, the reactor can be single-stage or multi-stage, and the column may be a simple volume or contain packing geometry to increase the gas hold-up and interfacial area.  

\begin{figure}[h]
    \centering
    \includegraphics[width=1\linewidth]{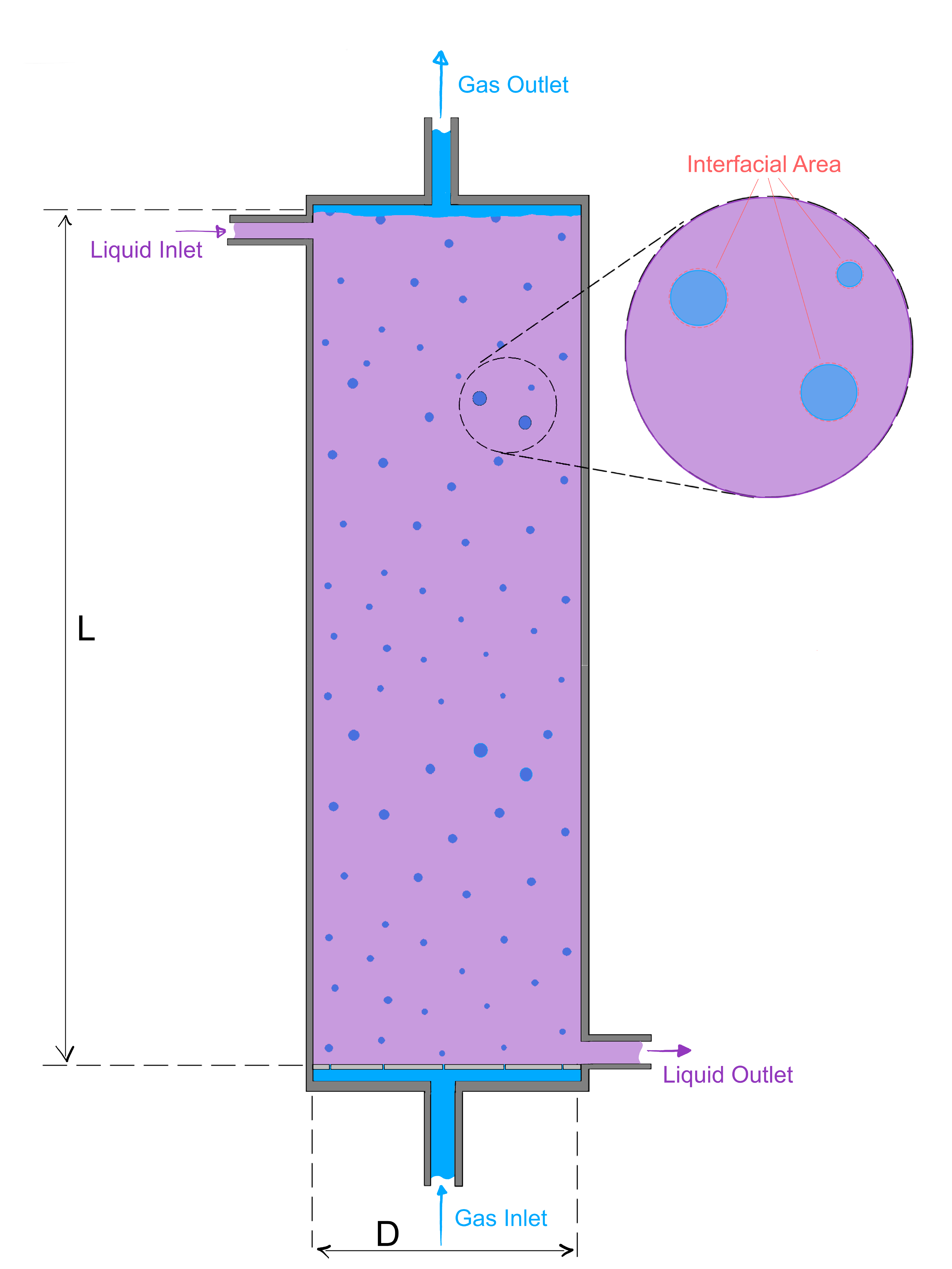}
    \caption{Sketch of a simple single-stage countercurrent Bubble Column Reactor GLC.}
    \label{fig:BCR sketch}
\end{figure}

In the following section, we describe a model for a simple (i.e. no packing), single-stage, countercurrent BCR (as illustrated in Figure \ref{fig:BCR sketch}).
\newline

\subsection{Mathematical model}
We take the BCR model for tritium extraction from liquid PbLi described by Malara \cite{malara_tritium_1995}, which defines the concentration of tritium in the liquid phase $c_T$ and the molar fraction of Tritium in the gas phase $y_{T_2}$ as a pair of second-order, 1D steady-state mass balance ODEs coupled by the term $J_{T}$:

\begin{equation}
\begin{gathered}
\text{Liquid Phase Mass Balance:} \\
\underbrace{\epsilon_{l}E_{l}\frac{d^{2}c_{T}}{dz^{2}}}_{\text{Dispersion}} + \underbrace{u_{l}\frac{dc_{T}}{dz}}_{\text{Advection}} - \underbrace{J_{T}a}_{\text{Mass Transfer}} = 0
\end{gathered}
\end{equation}

\begin{equation}
\begin{gathered}
\text{Gas Phase Mass Balance:} \\
\frac{1}{RT} \left( \underbrace{-\epsilon_{g}E_{g}\frac{d^{2}Py_{T_{2}}}{dz^{2}}}_{\text{Dispersion}} + \underbrace{\frac{d(u_{g}Py_{T_{2}})}{dz}}_{\text{Advection}} \right) - \underbrace{\frac{1}{2}J_{T}a}_{\text{Mass Transfer}} = 0
\end{gathered}
\end{equation}

Where $c_{T}$ is the liquid phase concentration of Tritium atoms [\si{mol/m^{3}}], $y_{T_{2}}$ is the molar fraction of $\ce{T_2}$ molecules in the gas phase, $P$ is the pressure [\si{Pa}], respectively for the gas and liquid phases $\epsilon_g$ and $\epsilon_l$ are the "hold-up" fractions, $E_g$ and $E_l$ are the axial dispersion coefficients, and $u_g$ and $u_l$ are the superficial velocities \si{[m/s]}, $a$ is the specific interfacial area \si{[m^{-1}]}, $J_T$ is the liquid to gas mass transfer flux \si{[mol\cdot m^2 / s]}, and $z$ is the axial position along the column \si{[m]}. 
Although we recognize that BCR extraction models have been further developed since Malara's work was published, this model provides a good example of how a physics-informed model can be integrated into PathSim.
\newline

Liquid breeders generate Tritium in solution, either in atomic form in metals or in diatomic form (eg. \ce{T_{2}}, \ce{TF}, \ce{HT}...) in the case of molten salts.
The equilibrium partial pressure of molecular \ce{T_{2}} in a gas phase in contact with a LiPb breeder is given by Sievert's law: 

\begin{equation}
c_{T}=K_{s} \cdot p_{T_{2}}^{0.5}
\end{equation}

Assuming that the rate-limiting factor is the mass transfer through the liquid metal boundary layer, the flux of Tritium atoms across the liquid-gas interface is given by \cite{Ricapito_tritium_2011}:

\begin{equation}
J_{T} = h_{l}(c_{T} - K_{s}(Py_{T_{2}})^{0.5})
\end{equation}
Where $K_{s}$ is the Sieverts constant of tritium in LiPb \si{[mol.m^{3}.Pa^{0.5}]}, and $h_{l}$ is liquid boundary layer mass transfer coefficient in \si{[m/s]}.

To solve the ODEs, the equations are made nondimensional and rewritten as a set of four first-order ODEs. Four boundary conditions describing the BCR inlet and outlet conditions of the liquid and gas phases allow the model solution to be found iteratively with a boundary value problem solver.

\subsection{Boundary conditions}
The boundary conditions for the gas and liquid inlets and outlets are defined as "open" or "closed". These describe whether the high-dispersion hydrodynamic environment within the column is modelled to extend into the upstream and downstream system ("open") or if dispersion ceases at the column boundaries ("closed"). 

With these conditions, we can model an "Open-Closed" and a "Closed-Closed" configuration. The "Closed-Closed" configuration is known as a Danckwerts boundary condition and represents a typical BCR design where the inlet feeds are much narrower than the column, isolating the high-dispersion region within the BCR.   The full details of the adimensionalisation and boundary conditions are available in the Github repository.

\subsection{Results}
We are unable to reproduce the results given in the Malara paper. The paper omits correlations for some of the parameters, for example the surface tension and interfacial area; however, even setting the parameters to the values stated in the paper does not produce the same results.

We instead compare our results to a more recent study by Mohan \textit{et al} \cite{mohan_experimental_2010} that used Malara's model and extended it to support multiple boundary conditions (namely "Closed-Closed" and "Open-Closed"). We calculate the specific interfacial area $a$ by assuming that the bubbles are evenly distributed, spherical, and have a mean diameter $d_{b}$:

\begin{equation}
    a = \frac{6\cdot\epsilon_{g}}{d_{b}}
\end{equation}

\begin{figure}[h!]
     \centering
     \begin{subfigure}[b]{\linewidth}
         \centering
         \includegraphics[width=1\linewidth]{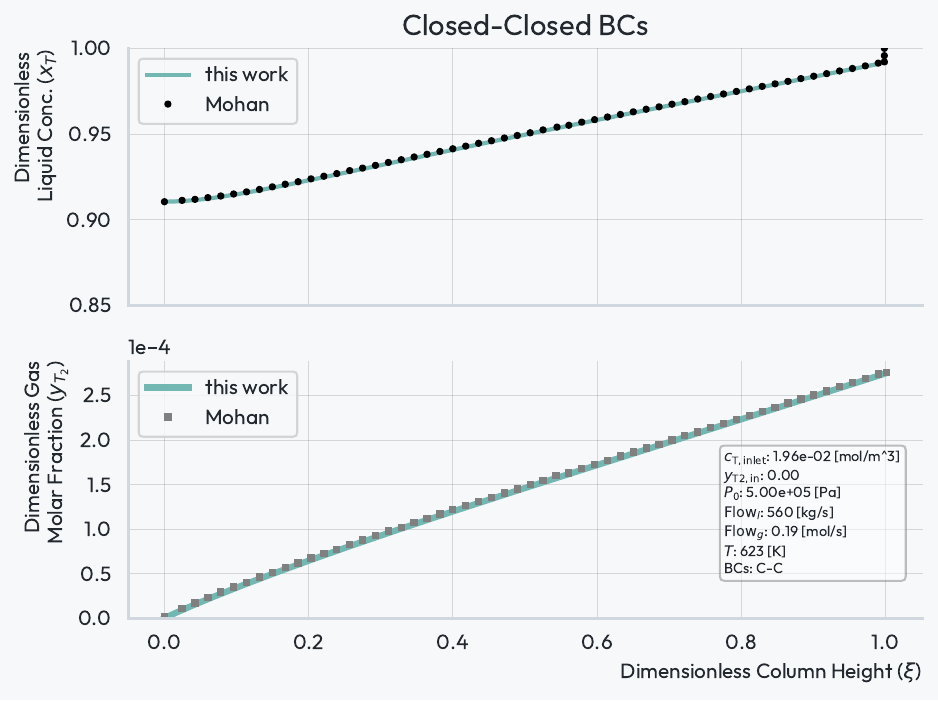}
         \caption{Closed-Closed case}
         \label{fig:mohan-cc}
     \end{subfigure}
     \begin{subfigure}[b]{\linewidth}
         \centering
         \includegraphics[width=1\linewidth]{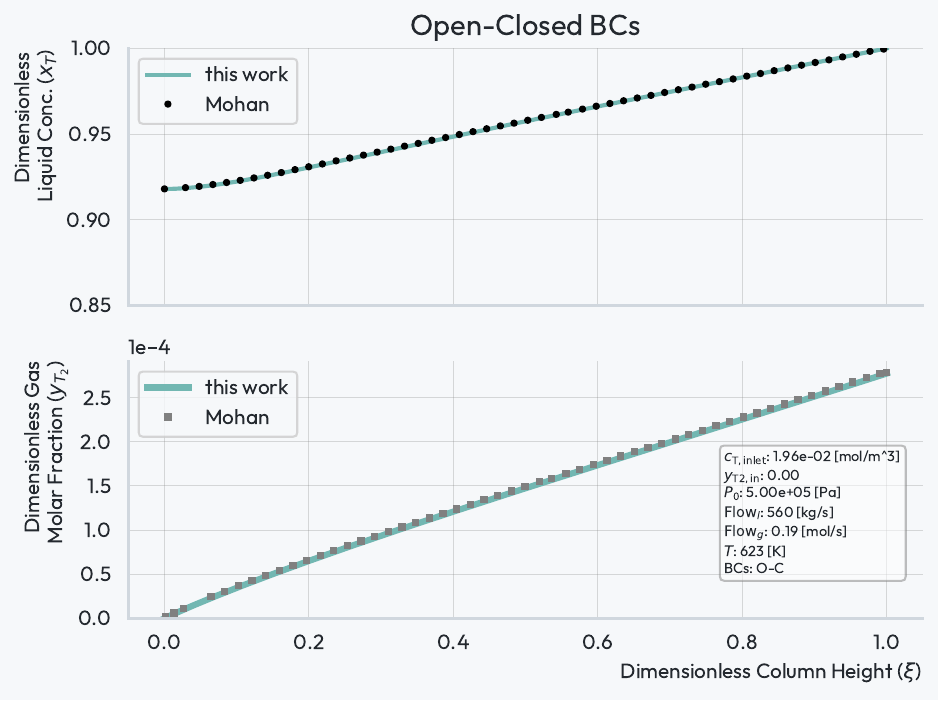}
         \caption{Open-Closed case}
         \label{fig:mohan-oc}
     \end{subfigure}
    \caption{Comparison of our solution of the BCR model vs previous work by Mohan \textit{et al} \cite{mohan_experimental_2010}.}
    \label{fig:comparison mohan}
\end{figure}

Using the correlations provided in the Malara paper for the bubble size $d_{b}$ and the volumetric mass transfer coefficient $ah_{l}$, we are able to reproduce exactly the results of the Mohan study (see Figure \ref{fig:comparison mohan}). This gives confidence that the model is implemented correctly.

This model can be used stand-alone (ie. not integrated with a dynamic system) to perform some dimensioning.
For example, we investigated the influence of the column diameter and height on the extraction efficiency.
The column diameter  and height  were varied between \SIrange{0.1}{1.0}{\metre} and \SIrange{1}{5}{\metre}, respectively.

\begin{figure}[h!]
     \centering
     \begin{subfigure}[b]{\linewidth}
         \centering
         \includegraphics[width=1\linewidth]{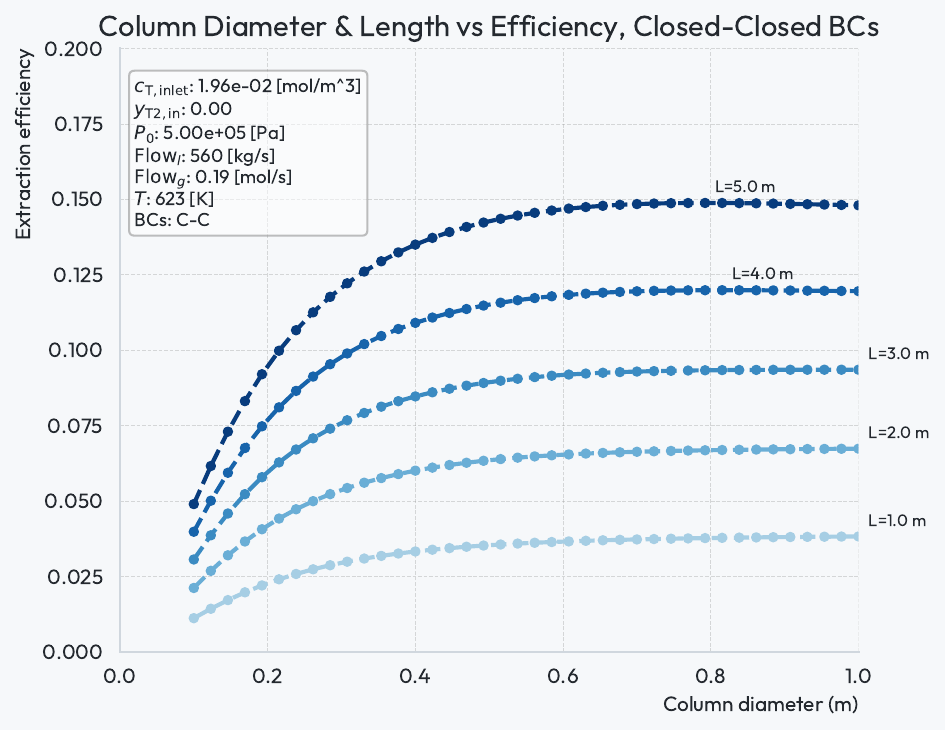}
         \caption{Closed-Closed case}
         \label{fig:column-parametric-scan-cc}
     \end{subfigure}
     \begin{subfigure}[b]{\linewidth}
         \centering
         \includegraphics[width=1\linewidth]{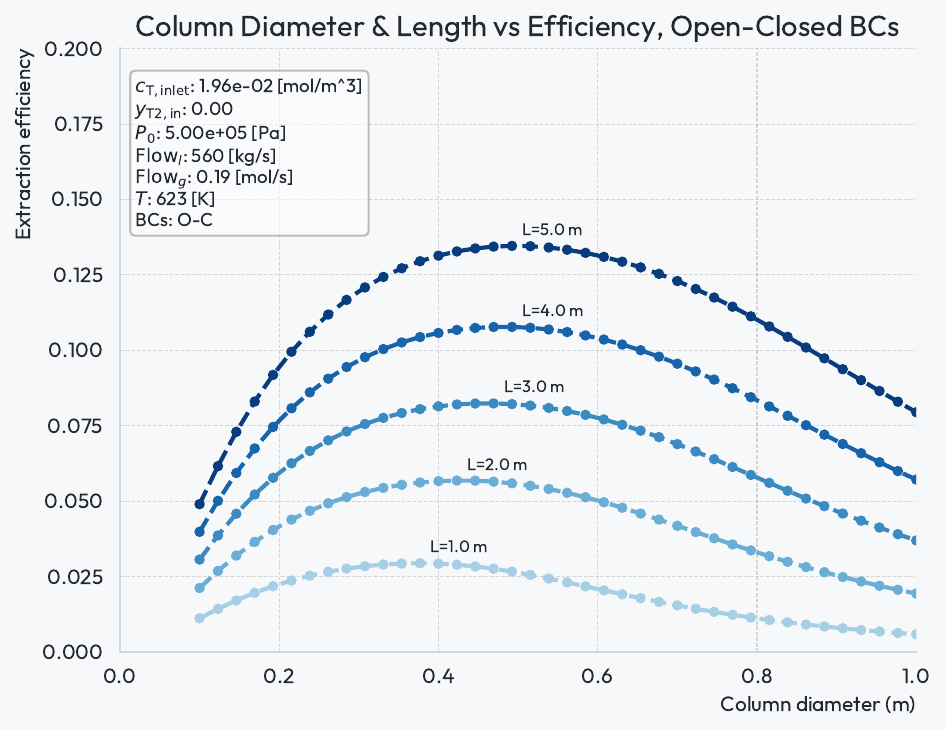}
         \caption{Open-Closed case}
         \label{fig:column-parametric-scan-oc}     \end{subfigure}
    \caption{Parametric scans of extraction efficiency vs bubble column height and diameter with O-C and C-C boundary conditions.}
    \label{fig:column-parametric-scan}
\end{figure}

When applying "Open-Closed" boundary conditions, the extraction efficiency first increases with the column diameter to reach a maximum before then decreasing (see \ref{fig:column-parametric-scan}). Increasing the diameter has competing effects. Initially, changes in the column's hydrodynamics enhance the mass transfer characteristics (e.g., gas holdup and interfacial area), which boosts efficiency. However, a larger diameter also significantly increases the axial dispersion. Under "Open-Closed" boundary conditions, this back-mixing flattens the concentration profile along the column, reducing the overall driving force for mass transfer. Above a certain diameter, this effect outweighs the gains from the improved mass transfer, and the efficiency starts to decrease.

This efficiency drop is not observed with "Closed-Closed" conditions. The formulation of the "closed" inlet boundary condition preserves the concentration gradient at the point of entry, which counteracts the internal back-mixing effects and instead causes the efficiency to plateau as the diameter increases.

As expected, the global extraction efficiency is highly dependent on the height of the column, ranging from a peak efficiency of just above \SI{2.5}{\percent} at $L=\SI{1}{m}$  to more than \SI{13}{\percent} at $L=\SI{5}{m}$ for the Open-Closed boundary condition with the given parameters.

\begin{table}[h] 
    \centering
    \caption{BCR Parameters used in Arc Fuel Cycle Models}
    \begin{tabular}{|l|c|c|} 
        \hline
        \textbf{Parameter} & \textbf{Value} & \textbf{Unit} \\ 
        \hline
        Temperature & 623 & °K \\
        Gas Inlet Pressure & 0.5 & MPa \\
        Column Height & 3 & m \\
        Column Width & 0.5 & m \\
        Total Liquid Flow & 560 & kg/s \\
        Total Gas Flow & 0.19 & mol/s \\
        Boundary Conditions & "C-C" & - \\
        \hline
    \end{tabular}
    \label{tab:bcr_data}
\end{table}
\subsection{Integration in a dynamic system}

\begin{figure}[h!]
     \centering
     \includegraphics[width=1\linewidth]{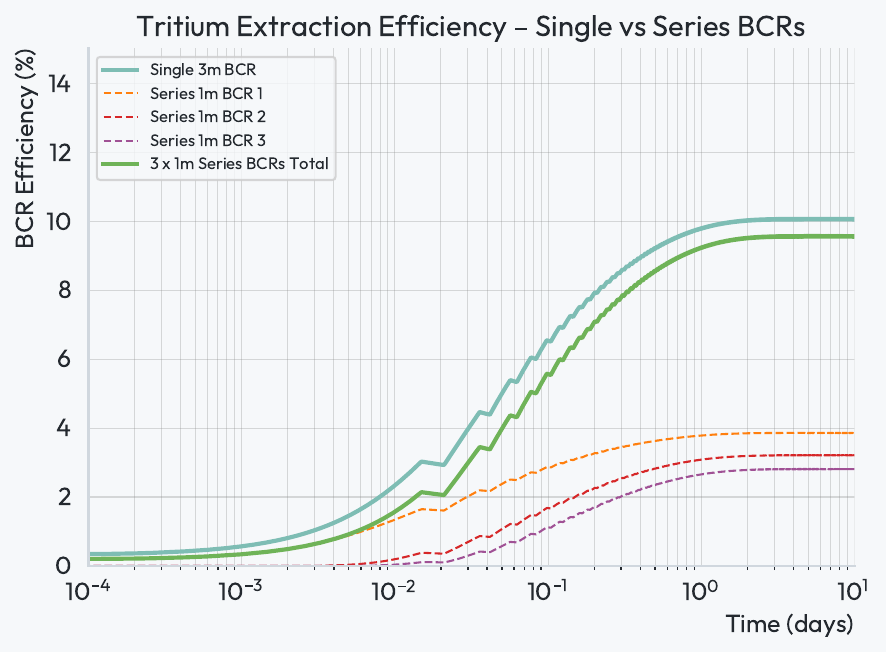}
    \caption{Comparison of the extraction efficiency of a single 3m high BCR vs 3 x 1m high columns in series, integrated as the TES in the 0D arc fuel cycle model.}
    \label{fig:BCR_efficiency_comparison}
\end{figure}

By placing the BCR model in a PathSim block, we can then use PathView to easily integrate this physics-informed model as the Tritium Extraction System (TES) in the previously described residence time model of the arc fuel cycle. This allows us to simulate the dynamic behaviour of the BCR, and enables comparison of different configurations of TES and their effect on the wider fuel cycle. Using the BCR parameters outlined in Table \ref{tab:bcr_data}, Figure \ref{fig:BCR_efficiency_comparison} shows a comparison of the extraction efficiency of a single 3m high column verses 3 x 1m high columns in series. In the series model, we assume that the gas is repressurised to the same inlet pressure between columns.
We can see that the steady-state total efficiency of the BCRs in series (\SI{9}{\percent}) is predicted to be slightly lower than a single BCR of the same total height (\SI{10}{\percent}). This is due to several effects; one obvious cause is the higher average hydrostatic pressure in the series columns, meaning that the average mass transfer driving force is lower.

\begin{figure}[h!]
     \centering
     \includegraphics[width=1\linewidth]{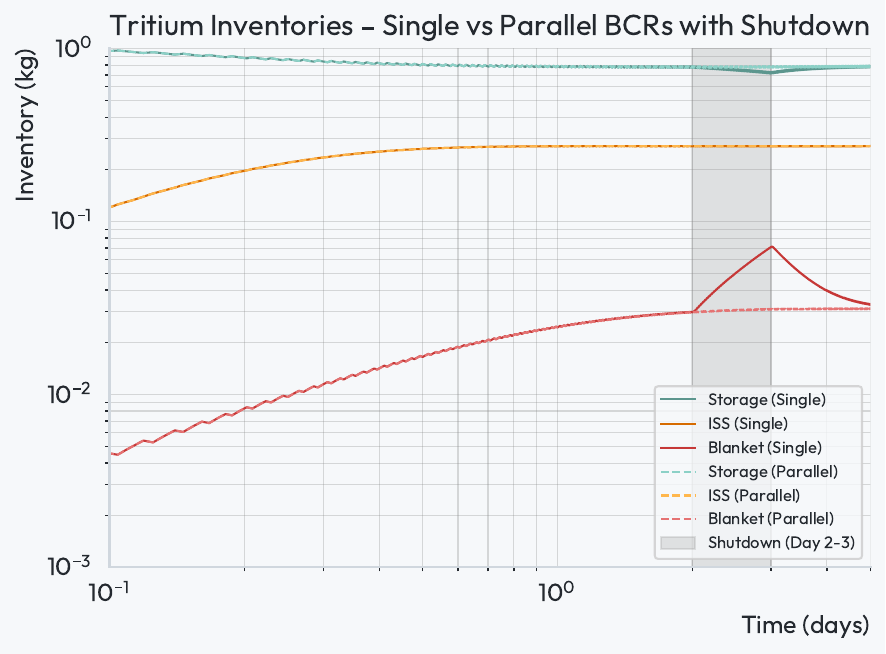}
    \caption{Comparison of the tritium inventories from 0 - 5 days for an arc fuel cycle with a single vs two parallel BCRs as the TES, with a shutdown of one BCR from day 2 - 3.}
    \label{fig:BCR_single_v_parallel_shutdown}
\end{figure}

With PathSim, we can also model events and study how these affect the wider system.
For example, using events we can simulate a period of downtime or maintenance on a BCR while fusion and tritium breeding continue. We can then compare the effect of BCR downtime on the fuel cycle for a TES with either a single BCR or two BCRs in parallel.
With a single BCR, tritium extraction ceases during downtime, inventory in the blanket accumulates and storage decreases (see Figure \ref{fig:BCR_single_v_parallel_shutdown}).
With two columns in parallel, the extraction efficiency actually improves marginally when one BCR is inoperative due to the increased mass flow through the remaining operational column.
There is no accumulation of blanket inventory during downtime, as the TES has redundancy with parallel BCRs.

\section{FESTIM models}

In certain situations, more detailed and complex numerical models are needed to simulate fuel cycle components and analytical models, residence time models, or simple finite difference models are not sufficient.
This is the case for instance when simulating tritium transport in a breeding blanket where tritium is bred from the interaction between lithium and neutrons \cite{dark_influence_2021} or tritium retention in plasma facing materials \cite{hodille_tritium_2026, delaporte-mathurin_festim_2024, delaporte-mathurin_3d_2023}.
PathSim allows us to "wrap" external tools in a dynamic model.
In this work, we will demonstrate this with the FESTIM code.

FESTIM is an open-source finite element framework for modelling the transport of hydrogen isotopes in materials \cite{dark_festim_2025}. 
It is a flexible and extensible tool for simulating diffusion, trapping, surface interactions, and other processes that govern hydrogen behaviour.

\subsection{Diffusion through a slab}

To demonstrate the integration of FESTIM models in PathView, we simulate a very simple problem: the diffusion of hydrogen through a slab.
In this 1D problem, the concentration is imposed on one side (the upstream side) $c = c_0$, while on the other side (downstream) the concentration is set as zero.

\begin{figure}[h!]
     \centering
     \begin{subfigure}[b]{\linewidth}
         \centering
         \includegraphics[width=1\linewidth]{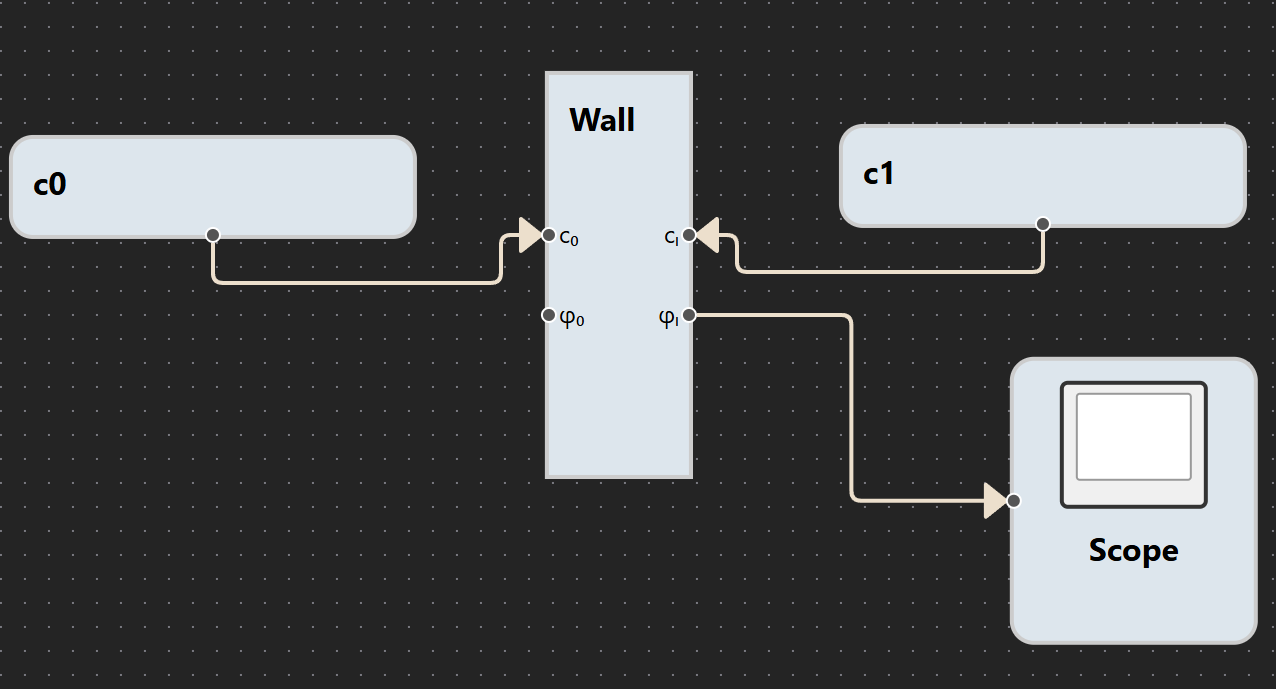}
         \caption{Pathview representation.}
         \label{fig:slab-diffusion-in-pathview}
     \end{subfigure}
     \begin{subfigure}[b]{\linewidth}
         \centering
         \includegraphics[width=1\linewidth]{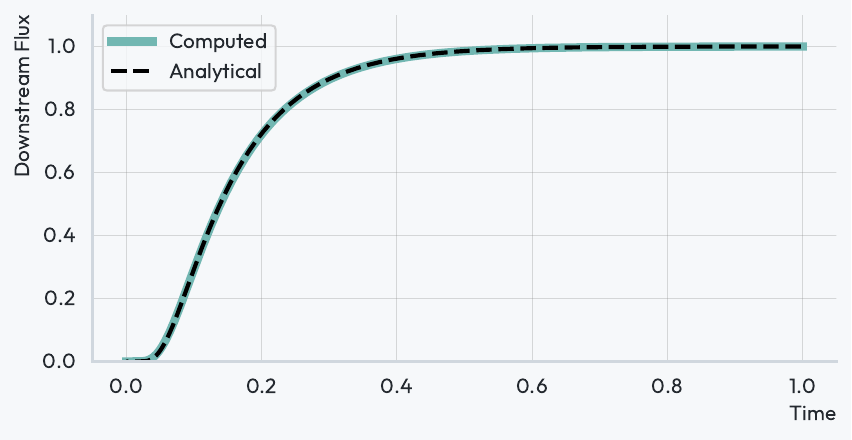}
         \caption{Comparison of the permeation flux simulated by FESTIM to the analytical solution.}
         \label{fig:slab-diffusion-results}
         \end{subfigure}
    \caption{Integration of the simple 1D diffusion problem in PathView.}
\end{figure}

The inputs of the model are the value of the imposed concentrations on both sides of the slab.
The outputs are the hydrogen fluxes on the boundaries (see \cref{fig:slab-diffusion-in-pathview}).

The analytical solution for the flux on the downstream side is:

\begin{equation}
    \varphi = \frac{c_0}{L} \left(1 + 2 \sum_{n=1}^{\infty} \left(-1\right)^{n} \exp{(- \frac{\pi^{2} D n^{2} t}{L^{2}})}\right) 
\end{equation}

where $L$ is the thickness of the slab, $D$ is the diffusion coefficient, and $t$ is the time.

There is a very good agreement between the simulated permeation flux and the analytical solution (see \cref{fig:slab-diffusion-results}).

While this is an extremely simple problem, this mechanism can be leverage to simulate several walls/slabs/barriers in series and have multiple FESTIM simulations connected together.

\subsection{Depleted source}

\begin{figure}[h!]
     \centering
     \begin{subfigure}[b]{\linewidth}
         \centering
         \includegraphics[width=1\linewidth]{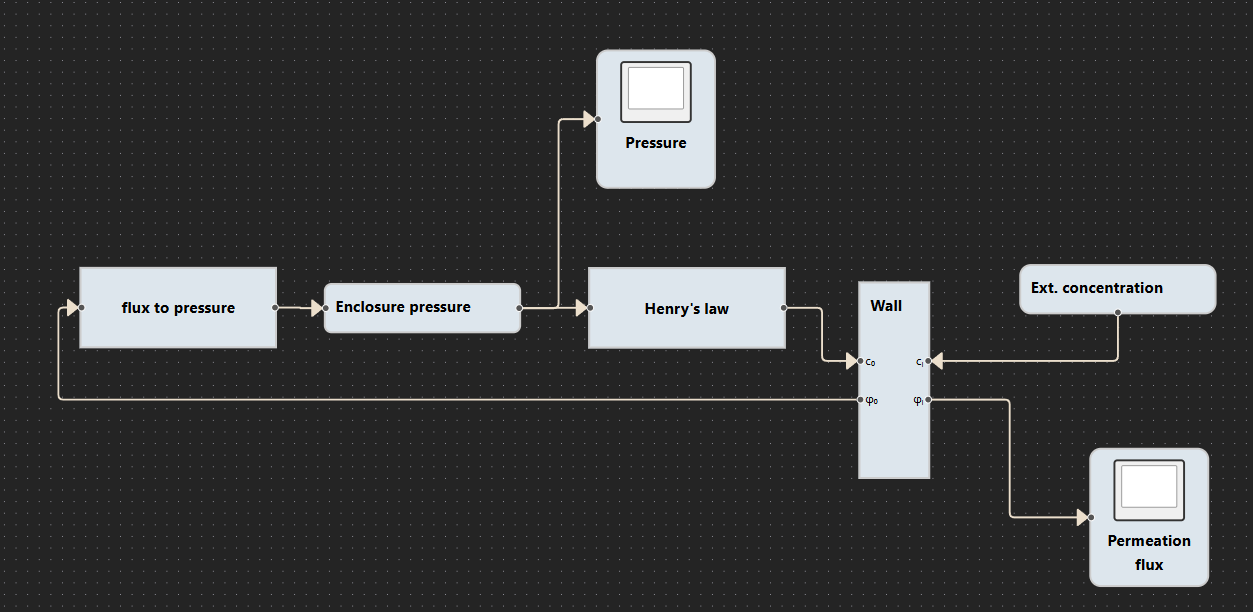}
         \caption{Pathview representation.}
         \label{fig:depleted-source-in-pathview}
     \end{subfigure}
     \begin{subfigure}[b]{\linewidth}
         \centering
         \includegraphics[width=1\linewidth]{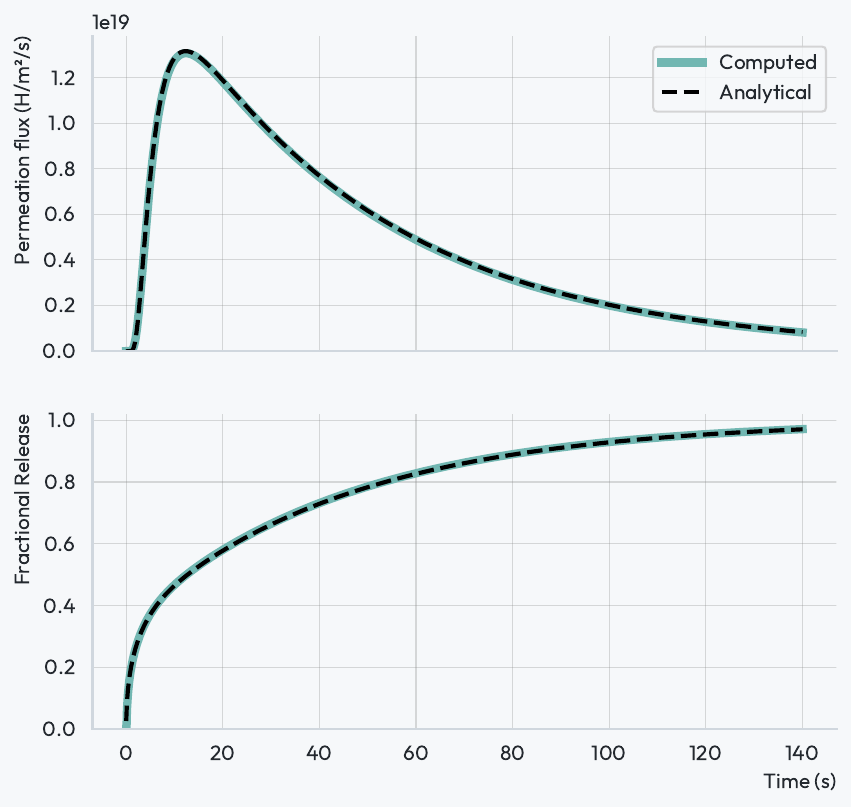}
    \caption{Comparison of the temporal evolution of the permeation flux (top) and the fractional release (bottom) from the enclosure simulated by FESTIM to the analytical solution.}
         \label{fig:depleted-source-results}
         \end{subfigure}
    \caption{Depleted source problem in PathView}
\end{figure}

This case from TMAP Verification and Validation report \cite{longhurst_verification_1992} consists of an enclosure containing an initial quantity of hydrogen gas (at an initial pressure).
This verification case was also added to the FESTIM V\&V book \cite{delaporte-mathurin_festim_vv_2024}.
The hydrogen can dissociate on the inner surface of the enclosure and then permeate through the walls.
As hydrogen particles escape the enclosure, the pressure decreases and as a result, so does the surface concentration on the inner walls.
This problem is, therefore, coupled and dynamic (see \cref{fig:depleted-source-in-pathview}).
At each time step, the flux of particles escaping the enclosure is computed, and the internal pressure is updated using the ideal gas law.

\section{Conclusion}

In this work, we have demonstrated three complementary levels of fidelity for modelling tritium fuel cycle components using the PathSim/PathView framework.

At the lowest level of detail, we employed the residence time method, which provides a computationally inexpensive, zero-dimensional representation of fuel cycle components.
This approach was used to reproduce the fuel cycle behaviour of an ARC-class fusion power plant, demonstrating its suitability for rapid system-level assessments and preliminary design studies.
The residence time model also served as a baseline upon which higher-fidelity component models could be progressively introduced.

The second modelling approach introduced physics-informed component representations based on coupled one-dimensional ordinary differential equations solved using finite difference methods.
This methodology was applied to model tritium mass transfer from liquid PbLi to a sparge gas in a bubble column reactor.
The standalone model was first validated through comparison with published literature before being embedded within the broader dynamic fuel cycle simulation, illustrating the seamless integration of intermediate-fidelity physics models.

Finally, we demonstrated the coupling of PathSim with the finite element tritium transport code FESTIM, enabled by their shared Python-based architecture.
This capability allows high-fidelity, multi-dimensional, multi-material, and multi-species tritium transport models to be incorporated directly into system-level fuel cycle simulations.
Such integration provides a pathway toward capturing spatial effects, material interfaces, and complex transport phenomena that cannot be represented using lower-fidelity approaches alone.

Overall, PathSim and PathView provide a flexible, open-source framework that enables fuel cycle modelling across multiple levels of physical fidelity within a single, coherent environment.
Future work will focus on extending this ecosystem through the integration of additional physics tools, such as OpenMC for neutronics, OpenFOAM for fluid dynamics, and surrogate models to accelerate high-fidelity component evaluations.
Additional physics-informed models will also be developed to simulate other fuel cycle components such as permeation against vacuum extractors, permeators, isotope separation systems, vacuum pumps, plasma facing components, breeding blankets, storage systems, etc.
By embracing an open-source development model, this framework aims to foster collaboration and accelerate progress across the fusion fuel cycle modelling community.

\section*{Data availability}

All scripts, models, datasets, and postprocessing are available on GitHub \url{https://github.com/rossmacdonald98/PathView_Paper}.

\section*{Acknowledgments}

The authors would like to acknowledge funding from the Advanced Research Projects Agency-Energy (ARPA-E), U.S. Department of Energy (DE-AR0001542).
The views and opinions of authors expressed herein do not necessarily state or reflect those of the United States Government or any agency thereof.
The authors would also like to acknowledge Prof. Samuele Meschini for allowing us to use and modify the ARC-class fuel cycle diagram figure.

\bibliographystyle{elsarticle-num} 
\bibliography{references}

\appendix 

\section{Additional figures}

\begin{figure*}[h!]
    \centering
    \includegraphics[width=0.9\linewidth]{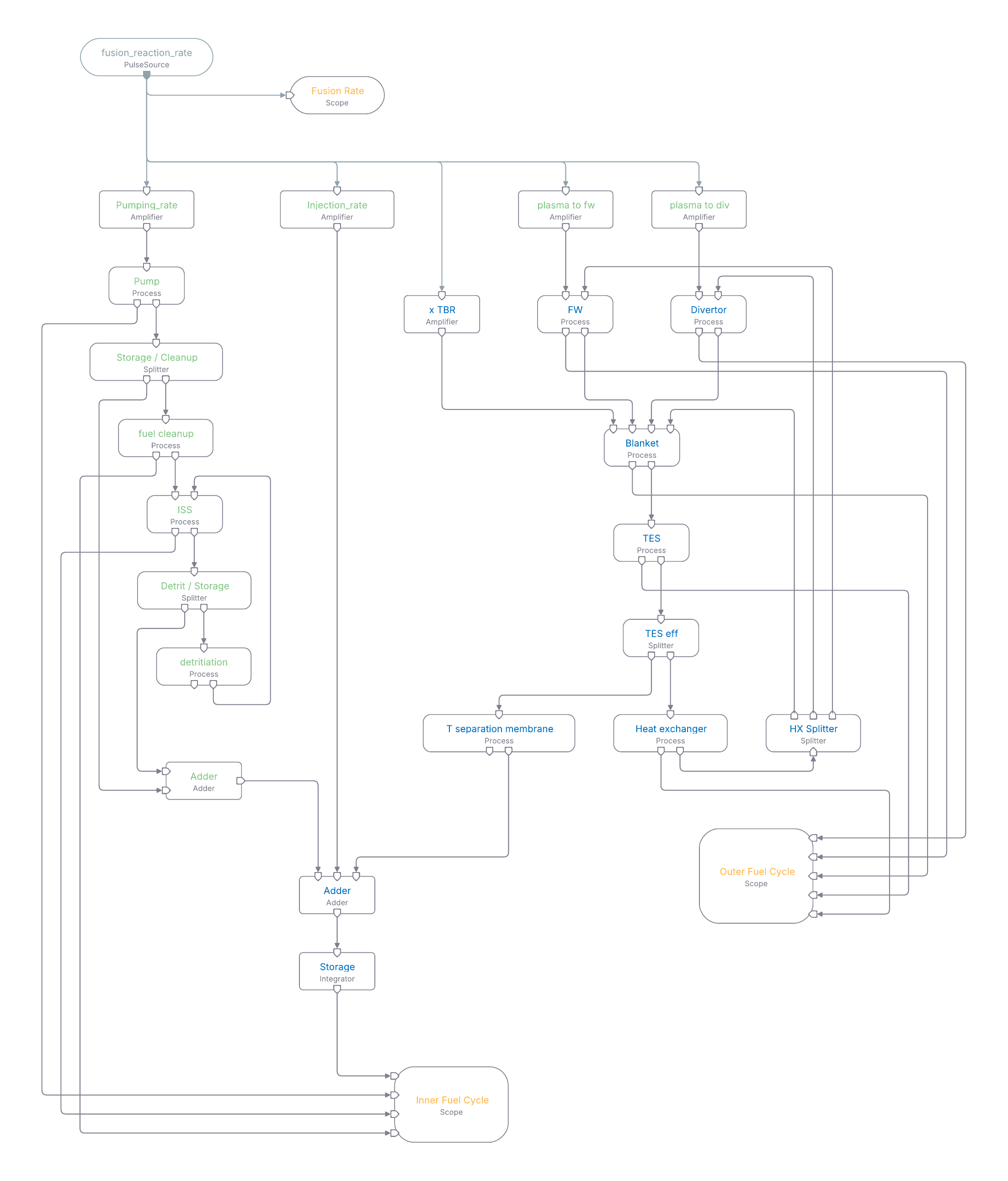}
    \caption{PathView graph of the ARC-class fuel cycle, replicating \cite{meschini_modeling_2023}.}
\end{figure*}

\begin{figure*}[h!]
    \centering
    \includegraphics[width=0.9\linewidth]{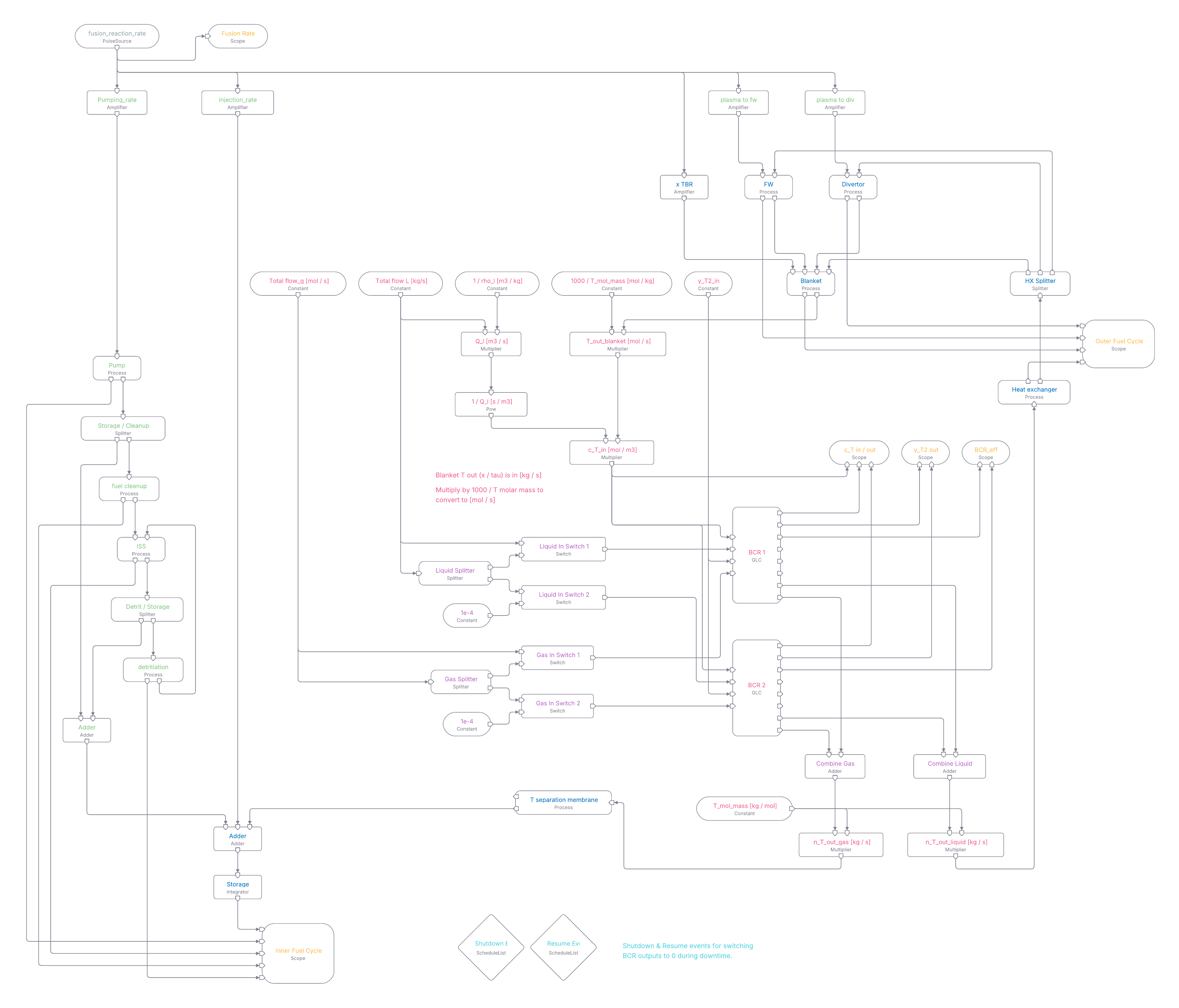}
    \caption{PathView graph of the ARC-class fuel cycle with parallel BCRs.}
\end{figure*}

\begin{figure*}[h!]
    \centering
    \includegraphics[width=0.9\linewidth]{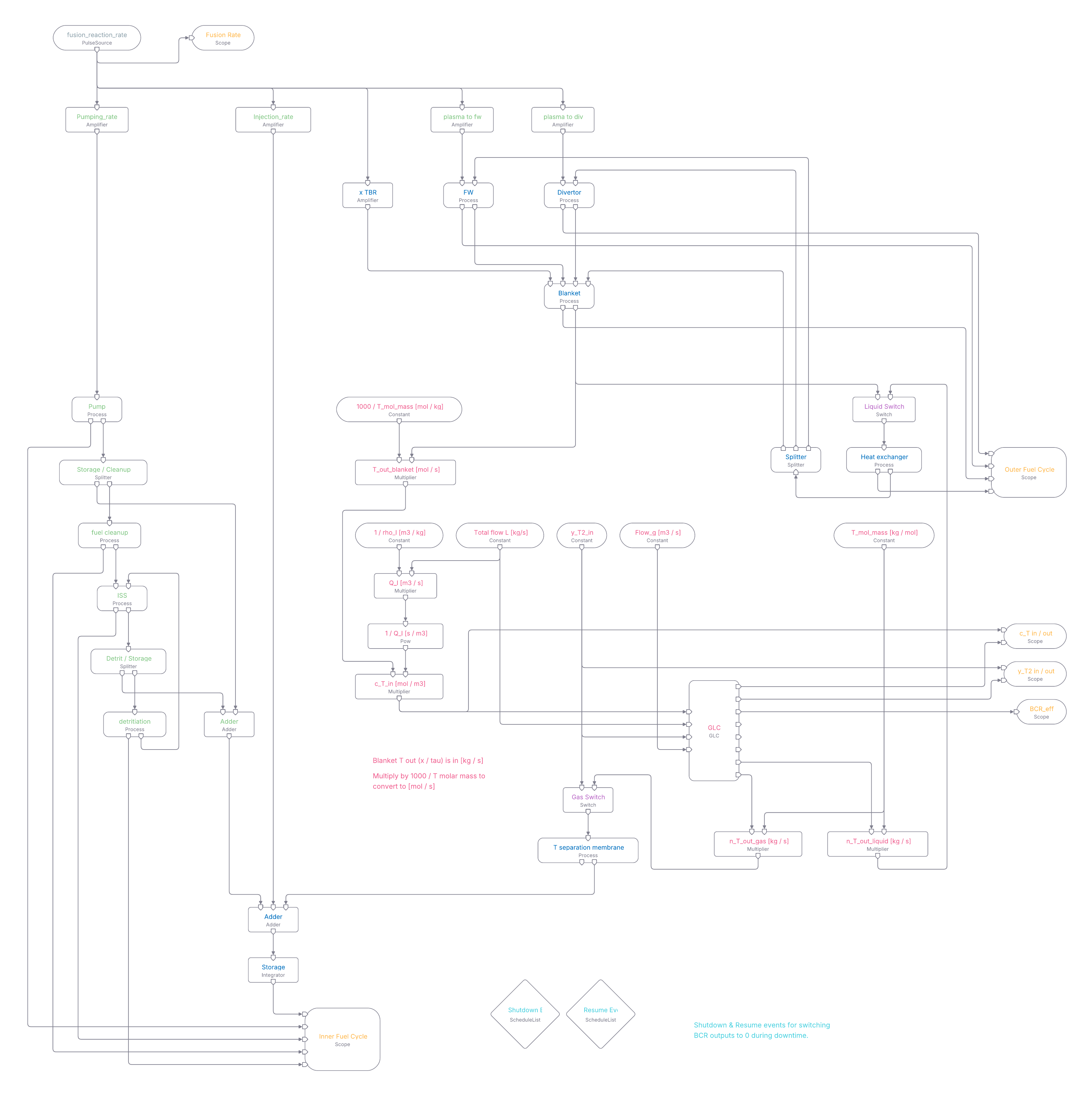}
    \caption{PathView graph of the ARC-class fuel cycle with a single BCR.}
\end{figure*}

\begin{figure*}[h!]
    \centering
    \includegraphics[width=0.9\linewidth]{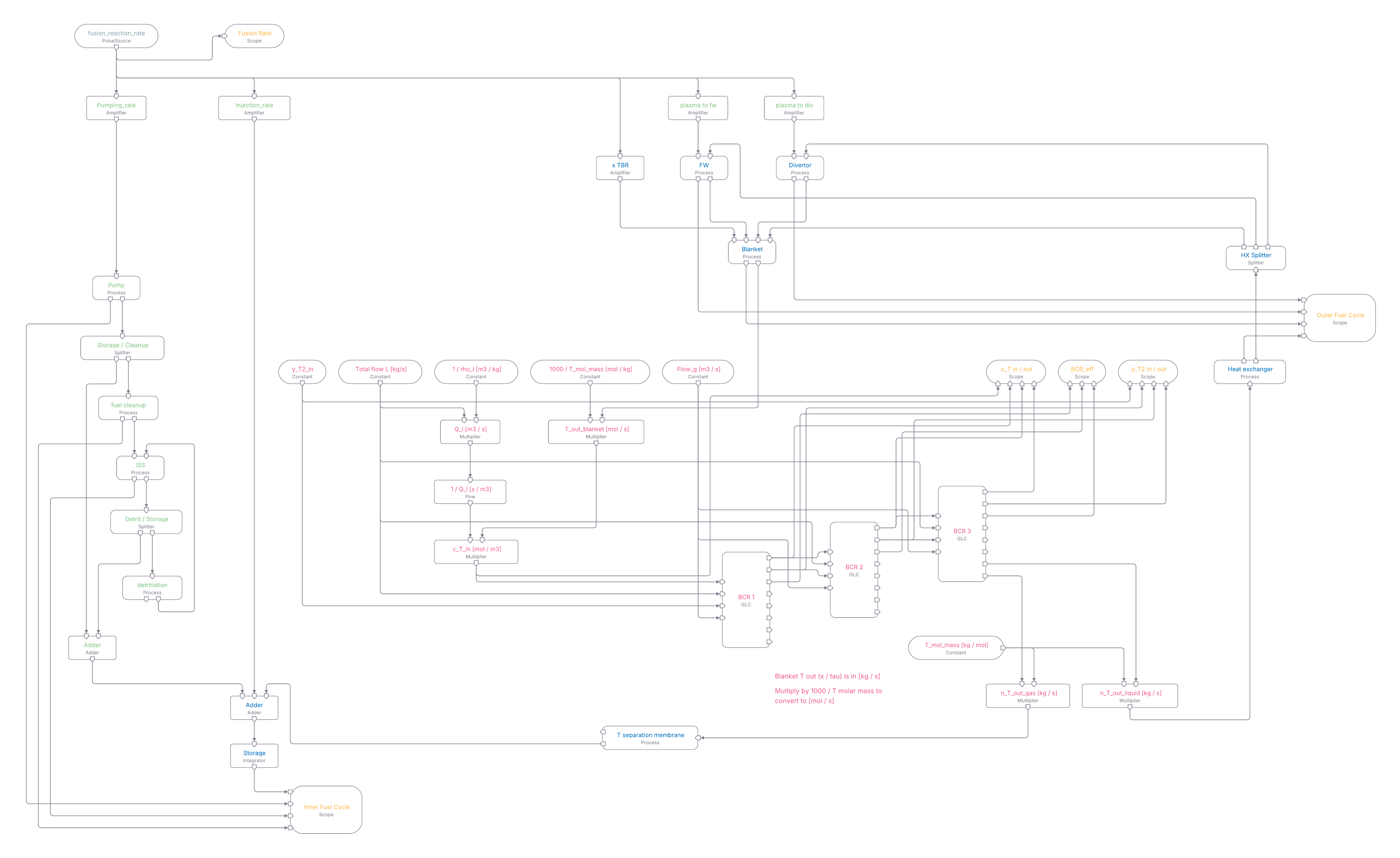}
    \caption{PathView graph of the ARC-class fuel cyclewith three BCRs in series.}
\end{figure*}

\end{document}